\documentclass[amssymb,amsmath,prl,twocolumn,floatfix,showpacs,superscriptaddress,numerical]{revtex4-1}

\usepackage{graphicx}
\usepackage[english]{babel}
\usepackage{subfig}
\usepackage{dsfont} 
\usepackage[colorlinks=true,allcolors=black]{hyperref}  

\newcommand{\ket}[1]{|#1 \rangle}

\begin{document}
\graphicspath{{pictures/}}
\title{Stimulated emission from NV centres in diamond}

\author{Jan Jeske}
\affiliation{Chemical and Quantum Physics, School of Sciences, RMIT University, Melbourne 3001, Australia}

\author{Desmond W. M. Lau}
\affiliation{ARC Centre of Excellence for Nanoscale BioPhotonics, School of Sciences, RMIT University, Melbourne, VIC 3001, Australia}

\author{Liam P.~McGuinness}
\affiliation{Institut f\"ur Quantenoptik, Universit\"at Ulm, Albert Einstein Allee 11, Ulm 89081, Germany}

\author{Philip Reineck}
\affiliation{ARC Centre of Excellence for Nanoscale BioPhotonics, School of Sciences, RMIT University, Melbourne, VIC 3001, Australia}

\author{Brett C.~Johnson}
\affiliation{School of Physics, University of Melbourne, Parkville 3010, Australia}

\author{Jeffrey~C.~McCallum}
\affiliation{School of Physics, University of Melbourne, Parkville 3010, Australia}

\author{Fedor Jelezko}
\affiliation{Institut f\"ur Quantenoptik, Universit\"at Ulm, Albert Einstein Allee 11, Ulm 89081, Germany}

\author{Thomas Volz}
\affiliation{ARC Centre of Excellence for Engineered Quantum Systems, Department of Physics and Astronomy, Macquarie University, North Ryde Sydney, NSW 2109, Australia}

\author{Jared H.~Cole}
\affiliation{Chemical and Quantum Physics, School of Sciences, RMIT University, Melbourne 3001, Australia}

\author{Brant C.~Gibson}
\affiliation{ARC Centre of Excellence for Nanoscale BioPhotonics, School of Sciences, RMIT University, Melbourne, VIC 3001, Australia}

\author{Andrew D.~Greentree}
\affiliation{Chemical and Quantum Physics, School of Sciences, RMIT University, Melbourne 3001, Australia}
\affiliation{ARC Centre of Excellence for Nanoscale BioPhotonics, School of Sciences, RMIT University, Melbourne, VIC 3001, Australia}

\email{janjeske@gmail.com}

\hyphenation{ab-sorp-tion}

\begin{abstract}
Stimulated emission is the process fundamental to laser operation, thereby producing coherent photon output.  Despite negatively-charged nitrogen-vacancy (NV$^-$) centres being discussed as a potential laser medium since the 1980's, there have been no definitive observations of stimulated emission from ensembles of NV$^-$ to date.  Reasons for this lack of demonstration include the short excited state lifetime and the occurrence of photo-ionisation to the neutral charge state by light around the zero-phonon line. Here we show both theoretical and experimental evidence for stimulated emission from NV$^-$ states using light in the phonon-sidebands.  Our system uses a continuous wave pump laser at 532 nm and a pulsed stimulating laser that is swept across the phononic sidebands of the NV$^-$. Optimal stimulated emission is demonstrated in the vicinity of the three-phonon line at 700 nm. Furthermore, we show the transition from stimulated emission to photoionisation as the stimulating laser wavelength is reduced from 700nm to 620 nm. While lasing at the zero-phonon line is suppressed by ionisation, our results open the possibility of diamond lasers based on NV centres, tuneable over the phonon-sideband. This broadens the applications of NV magnetometers from single centre nanoscale sensors to a new generation of ultra-precise ensemble laser sensors, which exploit the contrast and signal amplification of a lasing system.
\end{abstract}


\maketitle
\section{Introduction}
The negatively charged nitrogen vacancy centre (NV$^-$) in diamond \cite{Doherty2013} has found wide applicability in quantum information \cite{Aharonovich2009,Neumann2010, Henderson2011} and sensing \cite{Schirhagl2014} of magnetic fields \cite{Balasubramanian2008, Maze2008, Fang2013,Grinolds2013, Jensen2014, Rondin2014}, electric fields \cite{Dolde2011, Dolde2014}, temperature \cite{Neumann2013, Plakhotnik2014}, pressure\cite{Doherty2014} and biosensing \cite{McGuinness2011, Kucsko2013}. For both readout and sensing, fluorescence i.e.~spontaneous emission is collected and the spin state is inferred from the different emission levels. The time-averaged spontaneous emission from the different spin states differs by no more than 30\% and spontaneous emission is in all directions, creating the challenge of collecting the emission \cite{Clevenson2015}. 

Photons created by stimulated emission, on the other hand, have the same phase, wavelength and spatial mode as the stimulating light field. This enables close-to-perfect collection efficiency as all photons are emitted into the same mode. Furthermore, the competition between spontaneous emission and stimulated emission potentially allows contrast of several orders of magnitude between the brightest and least bright state ofthe NV$^-$ centre compared to the typical 30\% for single centres and 4\% for ensembles under spontaneous emission. 

Stimulated emission is also the basis for lasing, but despite the success of diamond Raman lasers \cite{Mildren2009} and despite an early proposal for a laser based on diamond defects \cite{Rand1986}, and the demonstration of cavity coupling \cite{Faraon2011, Jensen2014} a realisation of an NV laser has not been demonstrated so far. Microscopy, based on stimulated emission depletion (STED), a superresolution technique taking advantage of emission reduction by a strong light field, has been successfully implemented with NV centres \cite{Rittweger2009} as well as NVN centres \cite{Laporte2015}. However, the photophysical process leading to said emission reduction, be it stimulated emission to the ground state, or photo-ionisation to the charge neutral NV$^0$ state \cite{Manson2005, Waldherr2011, Beha2012} has not been extensively studied in this context. Indeed, photo-ionisation of the NV$^- \leftrightarrow$ NV$^0$ has been shown to be induced by strong light pulses\cite{Aslam2013, Siyushev2013, Chen2013, Shields2015}. The ionisation process has been suggested to be a two-photon process\cite{Waldherr2011, Aslam2013, Siyushev2013} which scales quadratically with laser power, for the wavelengths around the zero-phonon line (ZPL) of NV emission. Excited state absorption, for example via the ionisation channel, can lead to weakening of stimulated emission by driving towards NV$^0$. Particularly since both ionisation and stimulated emission by a strong light pulse result in reduced NV$^-$ spontaneous emission and are therefore hard to distinguish. 

Here we show experimentally that light with wavelengths above 650nm does not induce net ionisation for powers up to few W during the pulse) and show strong evidence for the presence of stimulated emission in the NV$^-$ phonon side band. We used a green (532 nm) CW pump-laser together with a pulsed laser source with tunable centre wavelength to measure the occurrence of stimulated emission at different wavelengths. The pulses, with a duration of only 6 ps, provide an intense light field and create a measurable change in the emission properties. Since the emitted photons are indistinguishable from the laser pulse, we detected the stimulated emission indirectly by the reduction of spontaneous emission. In addition, we measured the excited state population of both charge states separately by monitoring the corresponding ZPL intensities to detect any potential ionisation.

This paper is organised as follows: First we discuss stimulated emission using NV centres, our modelling results and describe the experimental setup to measure and maximise the effect. We then present the experimental spectra showing that the ratio of NV$^-$ and NV$^0$ emission is unaffected at the chosen stimulating wavelengths around 700 nm, followed by more detailed temporal measurements which further support stimulated emission. Finally we show that the effect occurs both in single-crystal diamond and nanodiamonds.

\begin{figure*}
\centering
\subfloat[]{\includegraphics[scale=1]{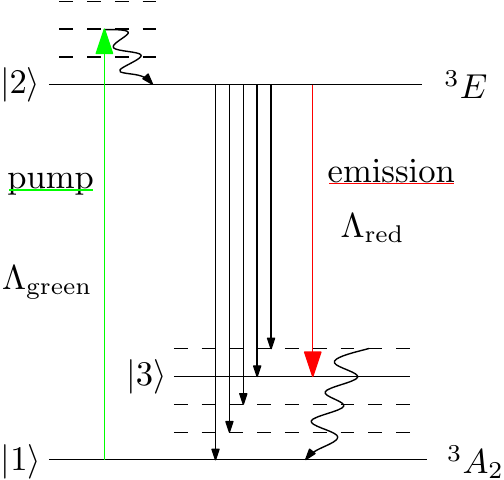}}
\hspace{1cm}
\subfloat[]{\includegraphics[width=\columnwidth]{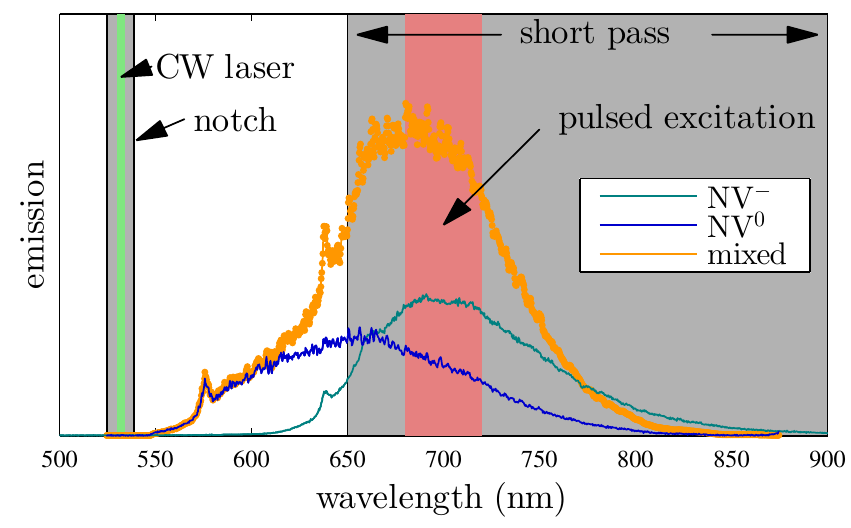}}
\hspace{1cm}
\subfloat[]{\includegraphics[scale=1]{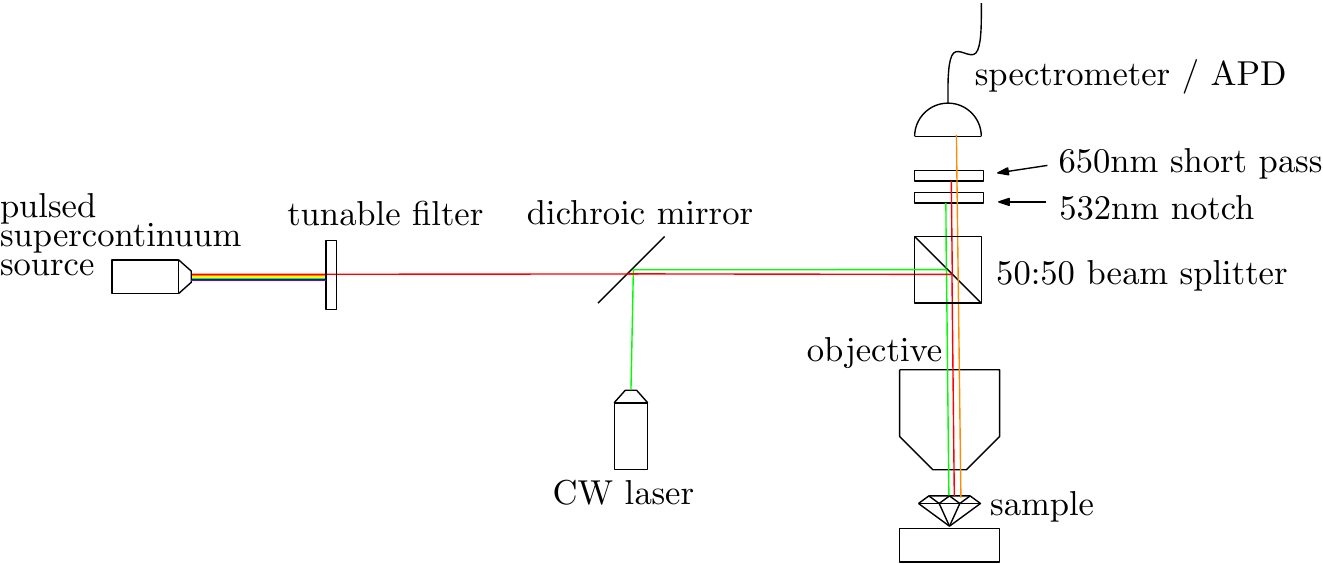}}
\caption{\textbf{(a)}: NV$^-$ level structure including phonon-added states (dashed and $\ket{3}$) with pumping and emission rates. Laser pumping (green) lifts population to the excited state, which can then decay via spontaneous emission (straight black arrows) or stimulated emission (red) due to interaction with a light field whose wavelength matches the $\ket{2} \leftrightarrow \ket{3}$ transition.
\textbf{(b)}: Laser excitation (red and green columns) with short pass and notch (grey). This setup blocks laser light from the detector but still allows to monitor both NV$^-$ and NV$^0$ emission. Example spectra are included: NV$^-$ emission (cyan), NV$^0$ emission (blue) and mixed emission (orange).
\textbf{(c)}: Experimental setup}
\label{fig setup}
\end{figure*}

\section{Stimulated emission and experimental setup}
Conventionally, both sensing and quantum state readout in the NV$^-$ centre is performed via fluorescence measurements after optical pumping, typically by exciting at a wavelength around 532 nm, from the $^3A_2$ ground state, which we denote $\ket{1}$, to the $^3E$ excited state $\ket{2}$, see figure~\ref{fig setup}(a). Spontaneous emission occurs at a rate $\approx(12$ ns$)^{-1}$ over the characteristic NV$^-$ spectrum with ZPL at 637 nm, corresponding to the $\ket{1} \leftrightarrow \ket{2}$ transition, and with phonon-sideband emission from 637 nm to $\sim$800 nm, corresponding to transitions from $\ket{2}$ to short-lived vibronic states slightly above $\ket{1}$ with added phonons, followed by a rapid decay to $\ket{1}$ \cite{Davies1976, Su2008}. 

Stimulated emission due to a coherent light field of wavelength $\lambda$ interacting with the NV centre occurs at a rate $\Lambda_{\rm{red}}=\sigma_\lambda \;P / (A \;\hbar \omega_\lambda)$ where $P/A$ is the power per area of the light field, $\hbar \omega_\lambda$ the corresponding photon energy and $\sigma_\lambda$ the stimulated emission cross-section, which as a function of $\lambda$ is proportional to the NV emission spectrum, i.e.~the spontaneous emission rate. Strong stimulated emission is achieved by strong fields, produced e.g.~in optical cavities or in our case by a pulsed supercontinuum source. 

The rate of stimulated emission is increased for laser light in the phonon-sidebands' wavelengths because the phonon-sideband emission is stronger than the ZPL emission, therefore the cross-section for decay on the phononic sidebands is higher. Accordingly we choose a pulsed laser excitation around 700 nm (red light) at the peak of the phonon-sideband, figure \ref{fig setup}(b), to achieve strong stimulated emission at the same wavelength. This laser is resonant with the transition between the excited state $\ket{2}$ and a phonon-added ground state $\ket{3}$ and induces a stimulated emission rate, which competes with the spontaneous emission, figure~\ref{fig setup}(a). The red laser will therefore enhance the emission probability around 700 nm and reduce the emission probability for the rest of the NV spectrum. It is also easier to achieve lasing on the sideband transition to the fast-decaying state $\ket{3}$ compared to the ZPL transition $\ket{1}$ since it is no longer necessary to pump more than 50\% of the NV centres out of the ground state $\ket{1}$. Our present setup however, is not designed for self-sustained lasing because the light field does not build up in a cavity but is provided by the pulsed laser. The stimulated emission can therefore not be measured directly, since it is indistinguishable from the stronger pulsed laser signal. However the reduction in the rest of the NV emission spectrum can be detected. This reduction can be turned on and off with the red laser. 

Obtaining full NV emission spectra is challenging when a strong laser at the centre of the NV emission spectrum ($\sim$700 nm) is used. We used a 650 nm short pass and a 532 nm notch in front of the detector to block out the pulsed red laser and the green (532 nm) CW pump laser respectively before the light was detected with a spectrometer or avalanche photodiode detector (APD), figure \ref{fig setup}(b). This detection window allowed us to monitor both the NV$^-$ ZPL at 637 nm as well as the NV$^0$ ZPL at 575 nm. Ionisation is expected to increase the emission of one peak at the cost of the other and can be clearly distinguished in this setup from stimulated emission, which should decrease the emission in the entire monitored range of wavelengths. Even in the case that ionisation NV$^-$ $\leftrightarrow$ NV$^0$ occurs both ways during the 6 ps pulse, the emission spectrum, which is mainly obtained in the 12.5 ns between pulses, should reflect the new relative charge state populations. The short pass optical filter allows us to vary the red laser wavelength i.e.~measure stimulated emission as a function of wavelength.

We used a pulsed supercontinuum source (Fianium WL SC400-8) with a 6 ps pulse duration at a repetition rate of up to 80 MHz with a tunable filter, figure~\ref{fig setup}(c), which allowed us to select both the centre wavelength (700 nm initially) and bandwidth (40 nm initially). The resulting time-averaged power of several mW corresponds to several Watts, during the pulse duration. The pulsed light was aligned via a dichroic mirror with the green CW laser and focused through an objective (NA=0.9) onto the NV diamond samples. We confirmed the effects on two $\sim$500~$\mu$m thick single-crystal diamonds as well as a nanodiamond sample, all with very high NV density. 

\begin{figure*}
\centering
\subfloat[]{\includegraphics[width=\columnwidth]{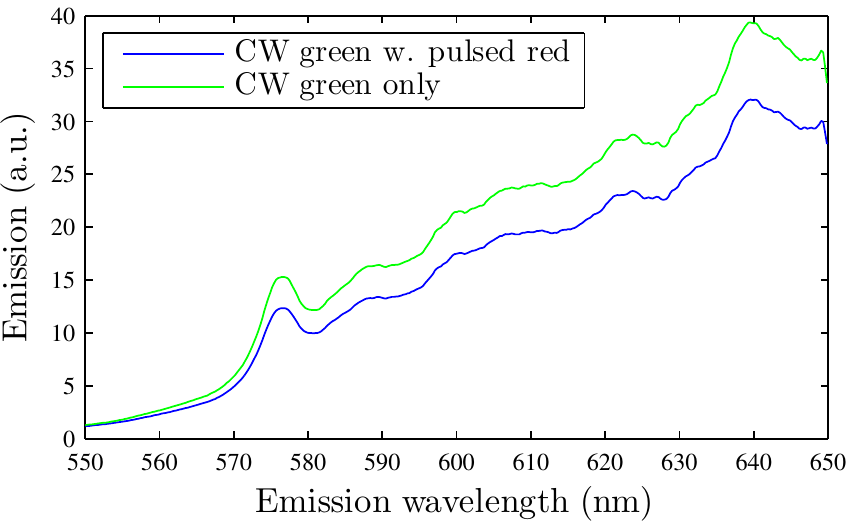}}
\subfloat[]{\includegraphics[width=\columnwidth]{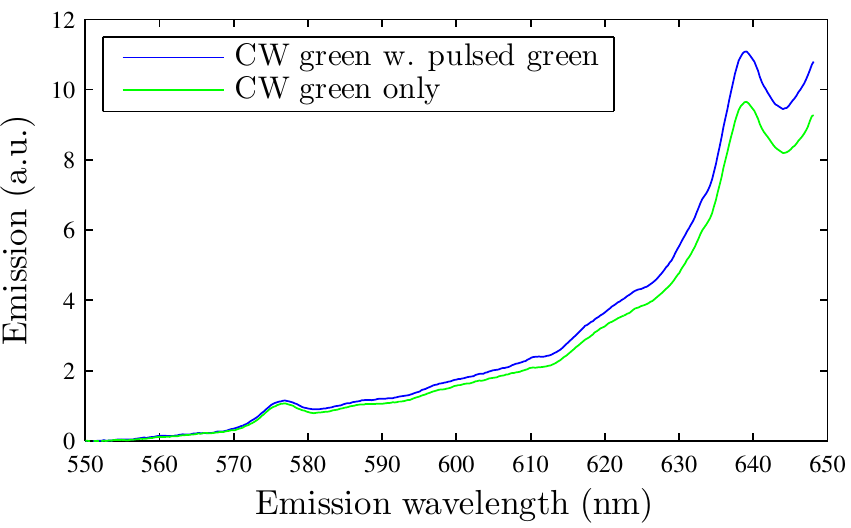}}\\
\subfloat[]{\includegraphics[width=\columnwidth]{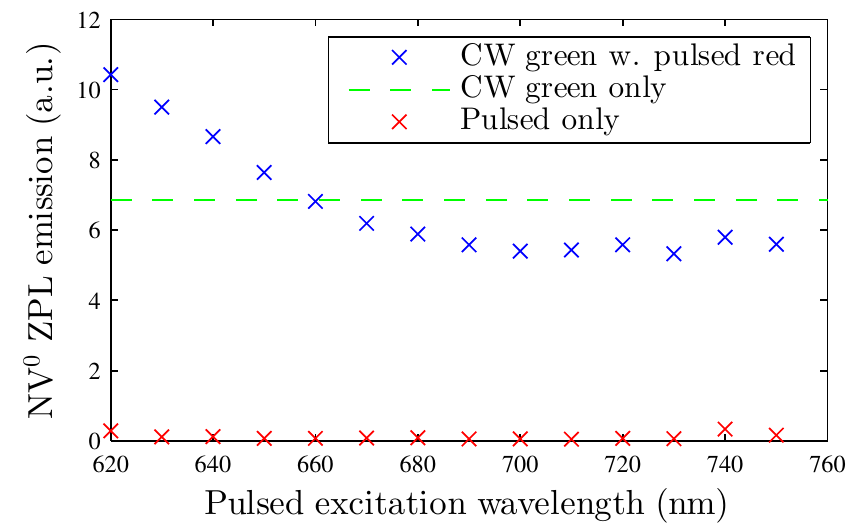}}\\
\caption{\textbf{(a)}: Spectral emission of the NV centre ensemble due to 532 nm CW laser excitation alone (green), and due to an added pulsed 700 nm laser excitation (blue). The emission with both lasers on is reduced compared to only green CW excitation. This is consistent with a preferential emission at 700 nm due to stimulated emission. \textbf{(b)}: When the pulsed laser wavelength is changed to also be at 532 nm the emission is increased by the pulsed laser. Both spectra were smoothed with a 4-nm-window. \textbf{(c)}  Two regimes exist: The NV$^0$ emission is decreased by the addition of pulsed laser wavelengths above 680 nm, where stimulated emission dominates, and increased for pulsed wavelengths below 640 nm, where ionisation from NV$^-$ to NV$^0$ is known to be the dominant process. In the measured range of wavelengths the pulsed laser alone does not create any NV$^0$ signal, i.e. there is no direct excitation. The pulsed power was kept constant at 2.9 mW for all excitation wavelengths.}
\label{fig spectra}
\end{figure*}

\section{Emission reduction of both ZPLs}
To investigate the effect of the pulsed laser on the spontanteous emission we first measured the NV emission between 550 nm and 650 nm when only the green CW laser excites the sample. We then measured  spectra when both the green CW and the pulsed laser at 700 nm with 40 nm bandwidth are on. The data in Fig.~\ref{fig spectra}(a) was taken over 100 ms acquisition time, which corresponds to 8 million pulses. The spectra are integrated over the entire acquisition time including the pulsed laser being on and off, and the spectra were smoothed with a running average over a 4-nm-window. Fig. \ref{fig spectra}(a) shows the emission with both the NV$^0$ ZPL at 575 nm and NV$^-$ ZPL at 637 nm clearly visible as expected since 532 nm pumping is known to excite both charge states of the NV. More importantly, the addition of the red pulsed laser \emph{reduces} the entire emission in the observed wavelengths as expected when stimulated emission at 700 nm is induced. We note that the reduction occurs despite the fact that the red pulsed laser by itself creates weak excitation of the NV$^-$ as discussed later. Both ZPLs (NV$^-$ and NV$^0$) are reduced, which indicates that the reduction of NV$^-$ emission is not caused by ionisation towards NV$^0$. If ionisation to NV$^0$ occurred, then we would expect the increased NV$^0$ population to provide an increase of the NV$^0$ ZPL emission. Quite contrary, in figure~\ref{fig spectra}(a) the NV$^0$ ZPL is also reduced indicating that the NV$^0$ population also experiences stimulated emission at 700 nm. While this might seem surprising, the phonon-sideband of the NV$^0$ extends over a slightly wider range than the NV$^-$ and is quite strong at 700 nm, see figure~\ref{fig setup}(b). 

Our interpretation attributes the emission reduction in the observed wavelength range to stimulated emission at 700 nm. As an initial check of this interpretation we changed the pulsed laser wavelength to also be 532 nm, as the CW source. This is outside the NV emission spectrum and stimulated emission should therefore not occur. Fig. \ref{fig spectra} (b) shows that in this case indeed the addition of the pulsed laser \emph{increases} the total emission, proving that the effect of emission reduction by the added pulsed laser only occurs at certain wavelengths. For stimulated emission this range of wavelengths would be expected to be the range of spontaneous emission, which we confirm further below. 

How can the apparent occurrence of stimulated emission be brought into agreement with the established fact \cite{Manson2005, Siyushev2013, Shields2015, Waldherr2011, Beha2012, Chen2013, Aslam2013} that strong light pulses around 637 nm efficiently decharge the NV$^-$ to the NV$^0$? To explore the photoionisation we reduced the pulsed laser wavelength, swapping the 650 nm short pass for a 600 nm short pass and  monitored the NV$^0$ ZPL. Again we combined green CW laser, which creates NV$^0$ emission, with the pulsed red light, which by itself does not create NV$^0$ emission. Any change in the NV$^0$ signal after adding pulsed light must therefore be due to either stimulated emission or ionisation. Fig.~\ref{fig spectra}(c) shows clearly the transition between two regimes: For pulsed light at high wavelengths $\lambda>680$ nm the NV$^0$ signal is reduced as discussed above and stimulated emission dominates. For pulsed light at lower wavelengths $\lambda<640$ nm the NV$^0$ signal is increased and ionisation from NV$^-$ towards the NV$^0$ is the dominant process \cite{Manson2005}. The threshold between the two behaviours seems to be around 660nm i.e.~a photon energy of 1.88~eV. This energy value is slightly higher than theoretical calculations of the minimal energy value required for ionisation from the NV$^-$ excited state, which were predicted as 1.56~eV in Ref.~\cite{Siyushev2013} and 1.7~eV in \footnote{private communications with Marcus Doherty}. Note that this value is not to be confused with the ionisation energy from the NV$^-$ ground state of 2.6~eV \cite{Aslam2013}.

The relative emission reduction at the NV$^-$ ZPL in Fig.~\ref{fig spectra}(a) is about 18\% and about 21\% in the measurement of Fig.~\ref{fig spectra}(c) for 700 nm pulsed wavelength. Assuming that the red intensity in the region of focus is intense enough to fully deplete the excited state via stimulated emission this means that 18 to 21\% of the NV centres which contribute to the signal are in the region of focus. Since we used a 500 $\mu$m thick sample the percentage value of focus volume relative to illuminated volume seems high but is realistic given that there is non-uniform intensity in all dimensions of the laser spot and considering that NV centres in the focus volume contribute more strongly to the emission signal.

\section{Emission dynamics}
While stimulated emission should only occur during the 6 ps pulse duration, a reduction by other indirect effects could show other time scales or no temporal structure. Changes in the emission due to ionisation have been found to occur typically on much slower time scales of $\mu$s to ms \cite{Manson2005, Waldherr2011, Chen2013}. 

\begin{figure*}
\centering
\subfloat[]{\includegraphics[width=\columnwidth]{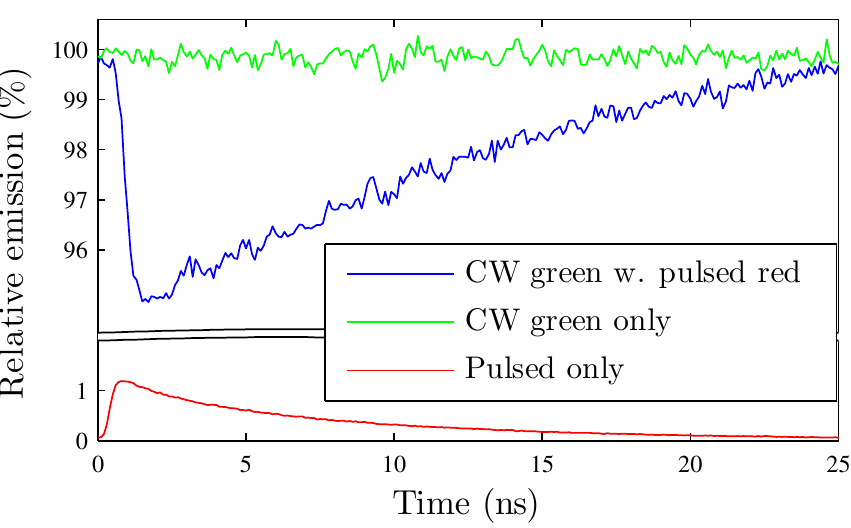}}
\subfloat[]{\includegraphics[width=\columnwidth]{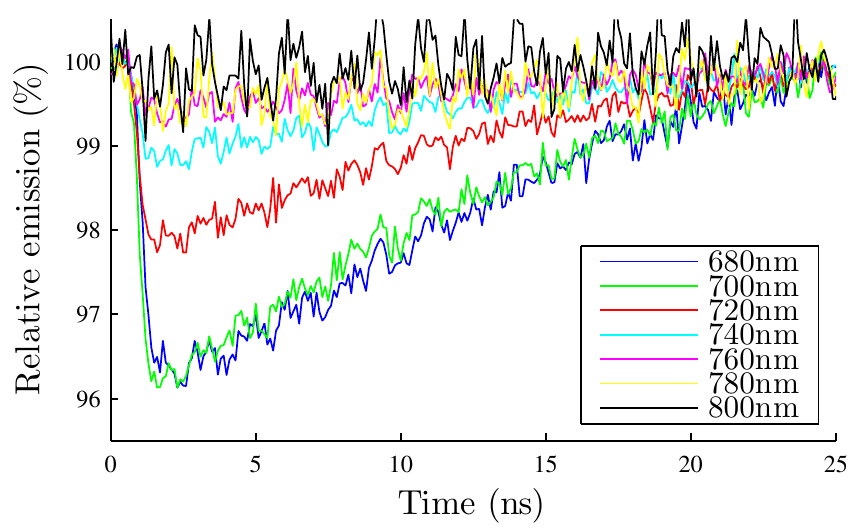}}\\
\subfloat[]{\includegraphics[width=\columnwidth]{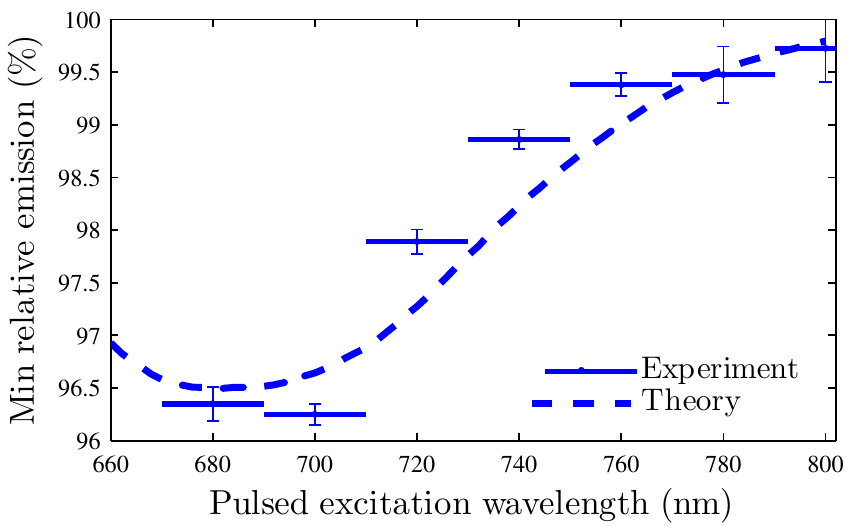}}
\caption{\textbf{(a)}: Raw data of temporally resolved emission which is spectrally integrated below 650nm. The fluorescence reduction occurs instantly at the point in time of the red pulse. Then the emission recovers in between pulses. \textbf{(b)}: Temporal emission curves for different pulsed excitation wavelengths \textbf{(c)}: The minimal relative emission directly after the pulse as a function of the pulsed excitation wavelength shows that the greatest reduction in emission occurs around 700 nm corresponding to the peak of the phonon-sideband of the NV.} 
\label{fig temporal}
\end{figure*}

We therefore measured the total emission (not spectrally resolved) in the detected emission window [550 nm to 650 nm, see Fig.~\ref{fig setup}(b)] with an APD and time-correlated counting card with a 0.1 ns temporal resolution. While the 6 ps pulse is not resolved on that scale, the dynamics between pulses is resolved. The time between pulses was set to 25 ns, corresponding to a repetition rate of 40 MHz. Fig.~\ref{fig temporal}(a) shows that the reduction due to the added pulsed laser (blue curve) relative to the green CW laser (green curve) occurs faster than the time-resolution of the electronics and then the signal recovers between pulses due to the green CW pump, as expected. The emission due to the pulsed laser only (red curve) shows that the 700 nm pulse alone actually weakly excites the NV$^-$ centre. The time point of this excitation coincides with the reduction of the emission of the blue curve, i.e. the time point of the pulse. 

To further investigate the wavelength dependence, discovered in figure~\ref{fig spectra}, we varied the centre wavelength of the pulsed laser in 20 nm steps. The full width at half maximum of the laser was  also set to 20 nm to ensure no overlap between the steps.  The strength of the reduction varies with the wavelength of the pulsed laser [figure~\ref{fig temporal}(b) and (c)] and shows that the effect is the strongest at 700 nm and reduces towards 800 nm. To ensure the data is independent of the small power variations of the source, which occurred when changing wavelengths, we measured the power for each wavelength and rescaled the reduction accordingly. Figure~\ref{fig temporal}(c) shows the relative emission level right after the pulse (averaged over 10 data points).  As expected for stimulated emission, the effect occurs for wavelengths within the NV emission spectrum and is strongest around 700 nm, which corresponds to the peak of the phonon-sideband. 

To understand the temporal dynamics we modelled the NV by a simple three state system [Fig.~\ref{fig setup}(a)]. This model neglects the spin dynamics but is a good approximation since the green CW laser spin-polarises into the $m_s=0$ state and all emission processes are spin-conserving. The differential equations for the system probabilities $P_j$ to be in the states $\ket{j}$ in Fig.~\ref{fig setup}(a) are:
\begin{align*}
\dot P_1 &= -\Lambda_{\rm{green}} P_1 + L_{21} P_2 + L_{31} P_3\\
\dot P_2 &= \Lambda_{\rm{green}} P_1 - [L_{21}+L_{23}+\Lambda_{\rm{red}}(t)]P_2 + \Lambda_{\rm{red}}(t) P_3\\
\dot P_3 &= [L_{23}+\Lambda_{\rm{red}}(t)] P_2 - [L_{31}+\Lambda_{\rm{red}}(t)] P_3
\end{align*}
where the pulsed laser created a stimulated emission rate $\Lambda_{\rm{red}}(t)$ with a temporal Gaussian profile with standard deviation 6 ps, repetition rate of (25 ns)$^{-1}=40$ MHz and overall strength is fitted to match the experiment. When we give time-independent values for $\Lambda_{\rm{red}}$ below, we give the equivalent rate that a 6 ps square pulse would have. The spontaneous emission rate to $\ket{3}$ was taken to be $L_{23}=18$ MHz from reference \cite{SuPhd}. All spontaneous emission rates to states other than $\ket{3}$ were grouped into one rate of $L_{21}=$65.3 MHz directly from state $\ket{2}$ to state $\ket{1}$. The value was chosen to correspond to a total lifetime of 12 ns for the excited state. The phonon-decay rate was set to $L_{31}=1$ THz but the exact value is irrelevant provided it is fast compared to all other transition rates. Starting the system in state $\ket{1}$ with the constant pumping rate $\Lambda_{\rm{green}}$ we time-evolved the system via numerical integration until it reached a steady state; we then included the time-dependent repeating red pulses in the evolution. 

In the simulations without the red pulses, the steady state has a given population in the excited state. Including the red pulses leads to stimulated emission on the $\ket{2}$ to $\ket{3}$ transition over the 6 ps pulse duration. This depletes population from the excited state to $\ket{3}$, which then decays to $\ket{1}$ due to the fast phononic decay. In the 25 ns between the pulses the population of the excited state slowly recovers to its original value due to the constant green pumping rate. The number of spontaneously emitted photons also drops during the pulse duration and then recovers in between pulses because it is proportional to the excited state population. The time scale of the recovery is roughly the time between pulses, i.e.~25 ns, and decreases (that is, the recovery rate increases) with increasing green CW pump power.

To quantify the expected wavelength dependence for the effect for stimulated emission we modelled different wavelengths of the pulsed laser by varying the stimulated emission cross section proportional to the measured NV emission spectrum with a maximum of $\Lambda_{\rm{red}}=6$ GHz at 682 nm. The pump rate was set to $\Lambda_{\rm{green}}=92$ MHz and figure \ref{fig temporal}(c) shows that the resulting theoretical wavelength dependence of the relative emission matches qualitatively with the experimental measurements. We note that a transition rate of order GHz is realistic for a pulsed source with powers of few W during the 6 ps pulse, and a focus-area on the order 1 $\mu$m$^2$.

The wavelength dependence is further evidence for stimulated emission as opposed to ionisation. Ionisation of the NV centre has been explained as an absorption process from the excited state to the conduction band. The absorption to the conduction band should therefore not show any wavelength dependence since the band can accommodate different photon energies. There should only be a sharp drop of the effect once the photon energy becomes too small (i.e. wavelength too long) to reach the conduction band. Ionisation should furthermore be more likely with increased excited state population, i.e. increased absorption. If the emission reduction was caused by ionisation then the wavelength dependence would follow the absorption spectrum of the NV, which has its phonon-sideband and maximum in wavelength range around 550 nm (below the ZPL) and not around 700 nm (above the ZPL) \cite{Manson2005}. 

While the pulsed laser wavelength changes the strength of the emission reduction, it does not seem to change the time scale of the recovery in figure~\ref{fig temporal}(b). This behaviour matches our simulations. The recovery of the emission in between pulses is approximately exponential. In the case that the effect is caused by stimulated emission the recovery should be purely caused by the green CW pump. For other effects, such as ionisation, the recovery could be either explained with the crystal returning to its equilibrium state, with a rate set by internal NV or crystal parameters which should be slower and independent of the green pump rate. For ionisation the recovery could also be explained by the green laser ionising back to the NV$^-$ state, which has previously been measured to occur at a rate of (0.25 $\mu$s)$^{-1}$ for a 532nm laser at 0.25 mW power \footnote{see Fig.~3(c) in \cite{Waldherr2011}} which is an order of magnitude slower than the rate we find.

\begin{figure*}
\centering
\subfloat[]{\includegraphics[width=\columnwidth]{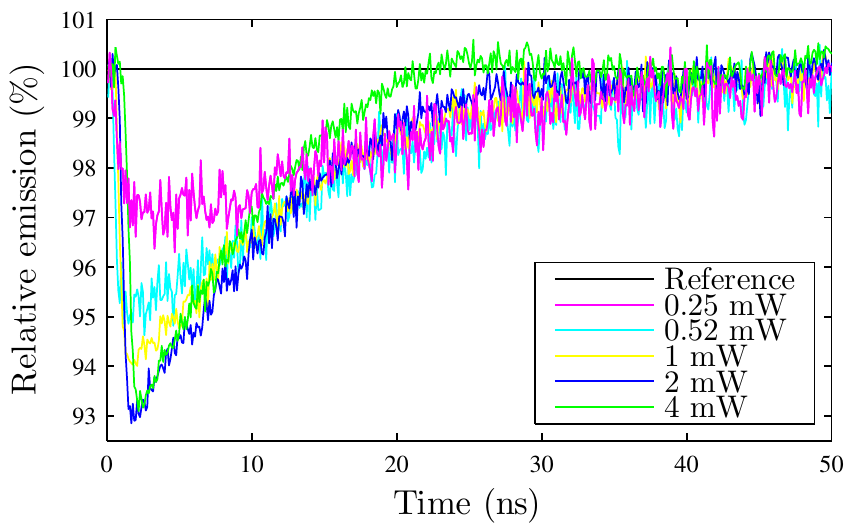}}
\subfloat[]{
  \includegraphics[width=\columnwidth]{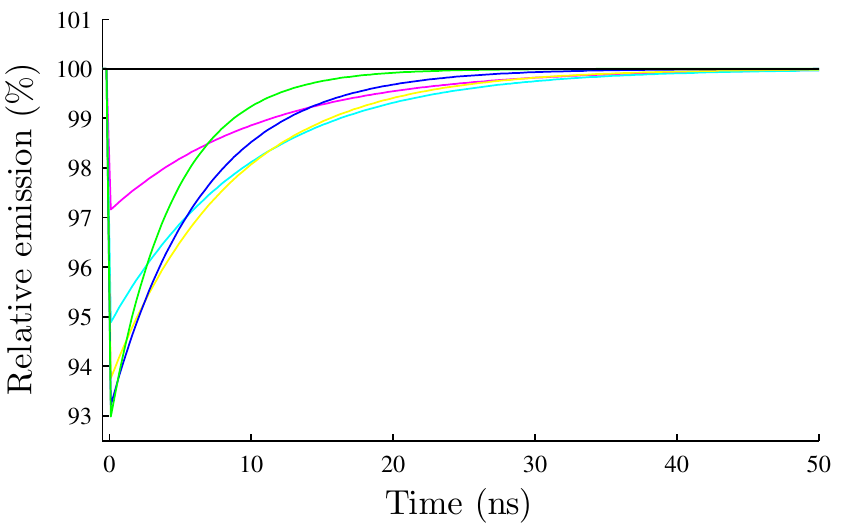}
  \llap{\raisebox{1.3cm}{
      \includegraphics[height=3cm]{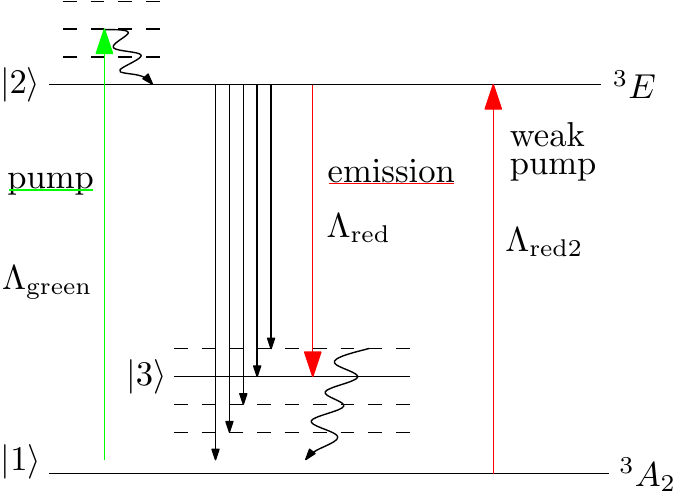}%
  }}
}

\caption{\textbf{(a)}: Emission reduction by the red pulse at $t\approx1$ns and recovery for different green CW powers. The recovery rate increases with increasing green power. The time-averaged red power was measured as 1.73mW. \textbf{(b):} The theoretical modelling, reproducing (a). }
\label{fig recovery rate - varying green power}
\end{figure*}

Figure~\ref{fig recovery rate - varying green power}(a) shows a measurement of the emission dynamics with varying powers of the green CW pump power. The pulse repetition rate was reduced to (100 ns)$^{-1}=$10 MHz to ensure that the system has enough time to return to its equilibrium state between pulses. Since stronger green CW pumping leads to higher total emission we plot the relative emission as a percentage of the highest total emission respectively. We observed a shorter recovery time, i.e. faster recovery rate, for stronger green CW power: In figure~\ref{fig recovery rate - varying green power}(a) the lowest power (pink curve) has the slowest recovery rate. With increasing green power (cyan, yellow, blue) the emission recovers faster and the highest power (green) shows the fastest recovery rate. The recovery time for 0.25 mW (pink) is $\sim$25 ns, which is much faster than the expected 250 ns for ionisation \cite{Waldherr2011}. The recovery time scales are consistent with stimulated emission as demonstrated below. For lower green powers the relative emission reduces to only $\sim$97\% (pink) rather than $\sim$93\% (green and blue). This is explained by the fact that the red pulsed laser weakly excites the NV as demonstrated below.

The observed dynamics was reproduced by our simulations: corresponding curves are plotted in the same colours in figure \ref{fig recovery rate - varying green power} (a) and (b). The stimulated emission rate was set to $\Lambda_{\rm{red}}=13$ GHz to reproduce the relative emission reduction to $\sim$93\% for large powers. The pump rates were varied proportional to the green laser powers in the experiment from 8.8 MHz to 141 MHz which corresponds to an average focus area of (0.6 $\mu$m)$^2$, given the known absorption cross-section $0.95\times 10^{-16}$ cm$^2$ \cite{Chapman2011}. These pump rates lead to recovery time scales similar to the experiment both in absolute values and relative trend between the curves. To reproduce the fact that lower green powers show a weaker emission reduction effect, we included an excitation rate of $\Lambda_{\rm{red 2}}=$0.85 GHz due to the red pulse from ground to excited state:
\begin{align*}
\dot P_1 &= -[\Lambda_{\rm{green}}+\Lambda_{\rm{red 2}}(t)] P_1 + L_{21} P_2 + L_{31} P_3\\
\dot P_2 &= [\Lambda_{\rm{green}}+\Lambda_{\rm{red2}}(t)] P_1 - [L_{21}+L_{23}+\Lambda_{\rm{red}}(t)]P_2\\ & \hspace{5.8cm}+ \Lambda_{\rm{red}}(t) P_3\\
\dot P_3 &= [L_{23}+\Lambda_{\rm{red}}(t)] P_2 - [L_{31}+\Lambda_{\rm{red}}(t)] P_3
\end{align*}

This rate $\Lambda_{\rm{red 2}}$ was calculated using the weak effective absorption cross-section for NV$^-$ at 700 nm, which we measured to be ${3\times 10^{-24}}$m$^2$. 
We had initially neglected this as it is only a weak effect but we found that for low green powers (which resulted in weak total emission) the inclusion of this effect in the modelling was a good explanation for why the pink curve only reduces to $\sim 97\%$ while higher green powers reduce to $\sim 93\%$.

\section{Nanodiamonds vs single crystal}
\begin{figure*}
\centering
\subfloat[]{\includegraphics[width=\columnwidth]{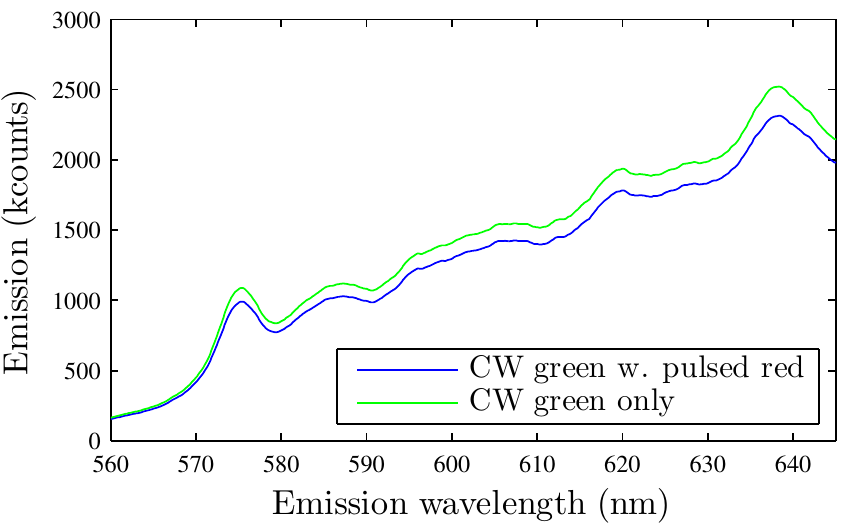}}
\subfloat[]{\includegraphics[width=\columnwidth]{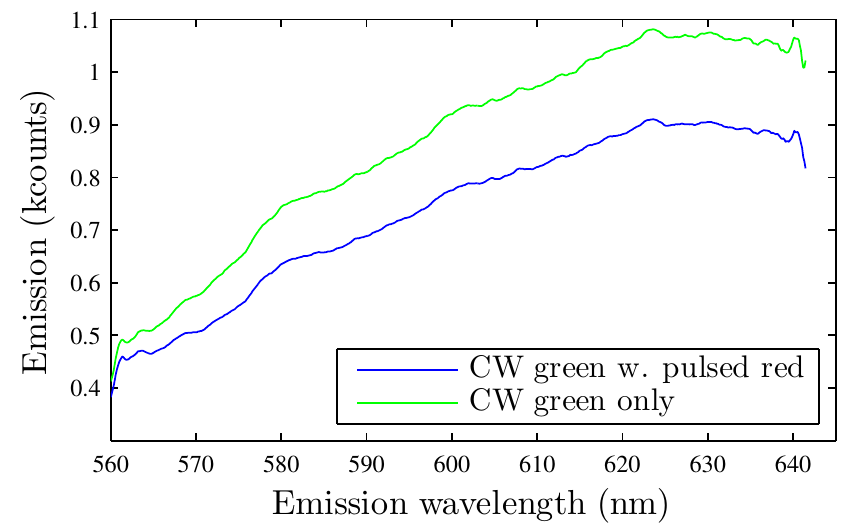}}\\
\subfloat[]{\includegraphics[width=\columnwidth]{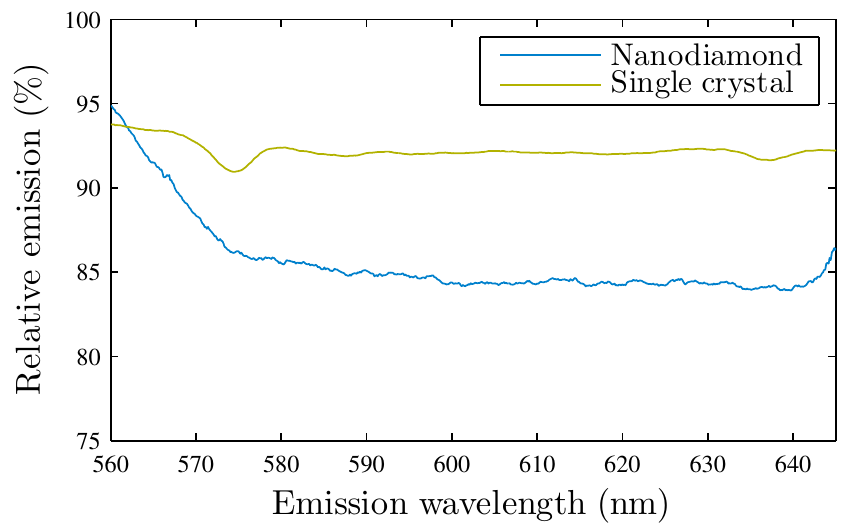}}
\caption{Comparison between single crystal (a) and nanodiamonds (b). The emission reduction by the pulsed supercontinuum source at 700 nm is seen for both samples. Dividing the lower curve by the upper curve we get the relative emission, shown in (c). The percentage reduction is stronger (although noisier) in the nanodiamond. The SC spectra were smoothed with a 4-nm-window, the noisier ND spectra were smoothed with a 13-nm-window.}
\label{fig nanodiamond}
\end{figure*}

All measurements so far were taken with single crystal diamond. To test the generality of the results we compared a single crystal sample to nanodiamonds on a silicon substrate under the same conditions. The effect of emission reduction in the single crystal [Fig~\ref{fig nanodiamond}(a)] could be reproduced with the nanodiamonds [Fig.~\ref{fig nanodiamond}(b)]. The ZPLs are washed out in the nanodiamond spectra which is not uncommon due to a larger variation of strain and other factors in the different nanodiamonds compared to the single crystal. The signal is also noisier in the nanodiamonds, which we counteracted with a larger smoothing-window of 13 nm compared to the 4-nm-window for the single crystal. The average reduction however is stronger in the nanodiamonds (to $\sim$85\%) than in the single crystal (to $\sim$92\%), see Fig.~\ref{fig nanodiamond}(c). This was expected because the single crystal has a thickness of approximately 500 $\mu$m and therefore a relatively large volume, some of which is out of focus. In this volume we expect the red laser to be too weak to create significant stimulated emission, but the green laser still strong enough to create background fluorescence. A high percentage reduction in the focussed volume is therefore likely combined with NV background fluorescence and low reduction from the defocussed volume. The nanodiamonds on the other hand had a size of approximately 120 nm and the majority of the illuminated volume was highly focussed. Uniform illumination over the sample is therefore much more likely for the nanodiamonds. Nanodiamonds are however less suitable to be used as a laser medium in an optical cavity because the large number of surfaces typically results in strong scattering and loss.

\section{Conclusions}
We have shown clear signatures of stimulated emission in NV centres in single crystal diamond as well as nanodiamonds. We demonstrated the spontaneous NV emission from a green CW pump laser and the reduction of this spontaneous emission due to a strong pulsed red laser at the centre of the phonon-sideband ($\sim$700 nm). We explain this reduction in spontaneous emission as being due to stimulated emission at 700 nm. The measured emission reduction follows all the expected behaviour for stimulated emission in terms of its dependence on the wavelength of the pulsed laser, the spectral reduction of both the NV$^-$ and the NV$^0$ ZPL, as well as the temporal dynamics, including emission reduction during the laser pulse and a recovery rate which varies with the green laser power. Our theoretical modelling of stimulated emission also reproduces the measured effects. 

Particularly we show that for wavelengths $\sim$700 nm stimulated emission is the dominant process over ionisation. This is indicated by no increase of either ZPL line, a wavelength dependence which follows the emission rather than the absorption spectrum and a temporal dynamics inconsistent with ionisation.  We furthermore showed the transition to the regime where ionisation dominates, which is when the pulsed laser is set to wavelengths around the NV$^-$ ZPL. Our results are therefore consistent with previously reported photoionisation studies NV$^- \leftrightarrow$ NV$^0$ induced by wavelengths below 650nm.

Our findings manifest that stimulated emission dominating over ionisation is dependent on the wavelength (rather than the intensity), which should be in the phonon-sideband $\lambda>650$ nm. This is in agreement with previous literature on superresolution microscopy, where stimulated emission depletion (STED) was achieved with wavelengths $\lambda>700$ nm pulses\cite{Han2009, Wildanger2011, Arroyo2013}, whereas charge state depletion was achieved with 637 nm pulses\cite{Han2010, Chen2015}.

The measurement of the effect in both single-crystal diamond and nanodiamonds shows sample-independent applicability.

Our results show that stimulated emission can be utilised in the NV and lead the way to exploring the benefits of stimulated emission, for example for higher collection efficiency into a cavity mode. Our results also indicate that NV centres are suitable as a laser medium. This opens up new avenues of applications for the NV centre. 

Laser threshold magnetometry \cite{Jeske2016LTM} was recently proposed as a novel way of measuring magnetic fields with NV centre ensembles, enabling higher contrast and stronger output signals using stimulated emission. By driving an NV laser at the threshold, sensitivities down to fT/$\sqrt(\rm{Hz})$ could be achieved and could enable a robust, room-temperature alternative for SQUID sensors. The current results support the feasibility of the laser threshold sensing scheme.

\acknowledgments
\section{Acknowledgements}
We thank Jason Twamley and Marcus Doherty for useful discussions and comments. We thank Takeshi Ohshima's group for their involvement in sample preparation. This work was supported by the Australian Research Council (ARC DP130104381).

\bibliography{aaaPublication}
\end{document}